\documentclass[pdftex]{article}
\usepackage{icrctc07}

\title{Ultra High Energy Cosmic Rays Diffusion in an Expanding Universe}
\shorttitle{UHECR Diffusion in an Expanding Universe}

\authors{R. Aloisio$^{1}$, V. Berezinsky$^{1,2}$, A. Gazizov$^{1,3}$ }
\shortauthors{Author and et al.}
\afiliations{$^1$INFN - Laboratori Nazionali del Gran Sasso, I-67010 Assergi (AQ), Itlay\\ 
$^2$Institute for Nuclear Research of RAS, 60th October Revolution prospect 7a, 117312 Moscow
Russia\\
$^3$Independence Avenue 68, BY-220072 Minsk, Republic of Belarus}
\email{roberto.aloisio@lngs.infn.it}

\abstract{We study the solution of the diffusion equation for Ultra-High Energy Cosmic Rays in the general case of an expanding universe, comparing it with the well known Syrovatsky solution obtained in the more restrictive case of a static universe. The formal comparison of the two solutions with all parameters being fixed identically reveals an appreciable discrepancy. This discrepancy is less important if in both models a different set of best-fit parameters is used.}

\begin{document}
\maketitle
%Begin the section.
\section{Introduction}
Diffusive propagation of Ultra High Energy Cosmic Rays (UHECR) in extragalactic space has been recently studied by \cite{AB04,AB05,Lemoine,Aloisioetal} using the Syrovatsky solution 
(see \cite{Syrov}) of the diffusion equation. The Syrovatsky solution is obtained under the 
restrictive assumptions of time-independent diffusion coefficient $(D=D(E))$ and energy losses of particles $(dE/dt=b(E))$. Recently two papers appeared \cite{BG1,BG2} solving the problem of the generalization of the diffusion equation (and its solution) in the case of an expanding universe, i.e. in  the case of time dependent diffusion coefficient and energy losses. In these works an analytic solution of the diffusion equation in an expanding universe was found, valid in the general case of time-dependent diffusion coefficient and energy losses, we will refer to this solution as the Berezinsky-Gazizov (BG) solution 
\cite{BG1}.  In the present paper we will compare, following the approach of \cite{BG2}, the spectra computed in the generalized case (BG solution) and the spectra obtained with the Syrovatsky solution as in the above cited papers. The  diffusion equation for ultra-relativistic particles propagating in an expanding universe from a single source, as obtained in \cite{BG1}, reads

$$ \frac{\partial n}{\partial t} - b(E,t)\frac{\partial n} {\partial
E}+3H(t)n - n\frac{\partial b(E,t)}{\partial E} - $$
\begin{equation} 
\frac{D(E,t)}{a^2(t)} \ \mathbf{\nabla}_x^2 n =
\frac{Q_s(E,t)}{a^3(t)} \ \delta^3(\vec{x}-\vec{x}_g),
\label{diff-basic}
\end{equation} 
where the coordinate $\vec{x}$ corresponds to the comoving distance and $a(t)$ is the scaling factor of the expanding universe, $n = n(t,\vec{x},E)$ is the particle number density per
unit energy in an expanding volume, $dE/dt =-b(E,t)$ describes the total energy losses, which include adiabatic $H(t)E$ as well as interaction $b_{int}(E,t)$ energy losses.
$Q_s(E,t)$ is the generation function, that gives the number of particles generated by a single source at coordinate $\vec{x}_g$ per unit energy and unit time. 

According to \cite{BG1}, the spherically-symmetric solution of Eq.~(\ref{diff-basic}) is

$$ n(x_g,E)=\int_0^{z_g} dz \left |\frac{dt}{dz}(z) \right | \ Q_s[E_g(E,z),z] \ $$
\begin{equation} 
\frac{\exp[-x_g^2/4\lambda(E,z)]}
{[4\pi\lambda(E,z)]^{3/2}} \ \frac{dE_g}{dE}(E,z),
\label{solution} 
\end{equation}
where
\begin{equation} 
\frac{dt}{dz}(z) = - \frac 1 {H_0 \ (1+z)\sqrt{\Omega_m(1+z)^3 +
\Lambda}}
\label{dtdz(z)} 
\end{equation} 
with  cosmological parameters $\Omega_m = 0.27$ and $\Lambda =0.73$,
\begin{equation} 
\lambda (E,z)= \int_0^z dz' \left |\frac{dt'}{dz'} \right | \
\frac{D(E_g,z')}{a^2(z')} , 
\label{lambda(E,z)} 
\end{equation} 
$$\frac{dE_g(E,z)}{dE} = (1+z)$$
\begin{equation} 
\exp\left [ \int_0^z dz'\left
|\frac{dt'}{dz'} \right | \ \frac{\partial
b_{int}(E_g,z')}{\partial E_g} \right ].
\label{dEg/dE} 
\end{equation} 
The generation energy $E_g=E_g(E,z)$ is the solution of the energy-losses equation:
\begin{equation}
\frac{dE_g}{dt}= - [ H(t)E_g + b_{int}(E_g,t)]
\label{dEdtB}
\end{equation}
with initial condition $E_g(E,0)=E$.

In the present paper we will discuss the propagation of UHE protons in Intergalactic 
Magnetic Fields (IMF) following the approach used by \cite{AB04,AB05}, in which the 
IMF is produced by a turbulent magnetized plasma. In this picture the IMF is characterized 
by a coherent field $B_c$ on scales $l>l_c$, where $l_c$ is the basic scale 
of turbulence, and on smaller scales $l<l_c$ the IMF is determined by its (assumed) turbulent spectrum. In our estimates we will keep $l_c\simeq 1$ Mpc. 

The propagation of UHE protons in IMF is characterized by two basic scales: an energy scale $E_c$ that follows from the condition $r_L(E_c)=l_c$, with $r_L$ Larmor radius of the proton,
and the  diffusion length $l_d(E)$, that is defined as the distance at which a proton is scattered by 1 rad. Using $l_d(E)$ the diffusion coefficient is defined as $D(E)=c l_d(E)/3$. 

We can easily identify two separate regimes in the particle propagation in IMF, that follows from the 
comparison of the two scale $r_L$ and $l_c$. In the case $r_L(E) \gg l_c$ ($E \gg E_c$) the diffusion length can be straightforwardly found from multiple scattering as
\begin{equation} %
l_d(E) = 1.2 \ \frac{E^2_{18}}{B_{\rm nG}} ~ {\rm Mpc},
\label{l_d} %
\end{equation} %
where $E_{18} = E/(10^{18}$~eV) and $B_{\rm nG}=B/(1$~nG). At
$E=E_c$,~ $l_d=l_c$. In the opposite scenario when $r_L<l_c$ $(E < E_c)$ the diffusion length depends on the IMF turbulent spectrum. In this case, following \cite{BG2}, we have assumed two different pictures: the Kol\-mo\-gorov spectrum $l_d (E) = l_c (E/E_c)^{1/3}$ and the Bohm spectrum 
$l_d(E) = l_c \ (E/E_c)$.

The strongest observational upper limit on the IMF in our picture is given by \cite{BBO99} as 
$B_c \leq 10$~nG on the turbulence scale $l_c = 10$~Mpc. In the calculations presented here 
we assume a typical value of $B_c$ in the range $(0.1 - 1)$~nG and $l_c = 1$~Mpc.

In the present paper we will not perform a detailed discussion of the proton diffusion in the general 
case of an expanding universe, we will address this issue in a forthcoming paper \cite{ABG07}, our 
main goal here is to perform a detailed comparison of the BG solution with the Syrovatsky solution. 
As already discussed in \cite{BG2},  the difference between these two solutions is substantial at energies $E \leq 3\times 10^{18}$~eV, where the effect  of the universe expansion (in
particular, of the CMB temperature growth with red-shift) is not negligible. The high energy tail 
of the UHECR spectrum is less affected by the expansion of the universe, nevertheless it is 
interesting to test the compatibility of the BG and Syrovatsky spectra at these energies where 
a substantial agreement of the two is expected. 

\section{Diffusive energy spectra of UHECR}

In the present calculations we used a simplified description of the IMF evolution with redshift, namely we parametrize the evolution of magnetic configuration $(l_c,B_c)$ as
$$ l_c(z) = l_c/(1+z), \ \ \   B_c(z)= B_c \ (1+z)^{2-m}, $$
where the term $(1+z)^2$ describes the depletion of the magnetic field with time due to the magnetic flux conservation and $(1+z)^{-m}$~ due to MHD amplification of the field. The critical energy 
$E_c(z)$ found from $r_L(E) = l_c(z)$ is given by
$$ E_c(z)=0.93 \times 10^{18}\ (1+z)^{1-m}\ \frac{B_c}{1~\mbox{nG}} $$
for $l_c = 1$~Mpc. The maximum redshift used in the calculations is $z_{\rm max}=4$.

Following \cite{AB05}, we have computed the diffuse flux assuming a distribution of sources on a lattice with spacing $d$ and an injection spectrum, equal for all sources, given by
\begin{equation} %
Q_s(E) = \frac{q_0 (\gamma_g - 2)}{E_0^2} \left( \frac{E}{E_0}
\right)^{-\gamma_g},
\end{equation} %
where $E_0$ is a normalizing energy (we used $E_0=1\times 10^{18}$ eV) and $q_0$ represents
the source luminosity in protons with energies $E \geq E_0$, $L_p(\geq E_0)$. The corresponding emissivity $\mathcal{L}_0 = q_0/d^3$, i.e. the energy production rate in particles with $E \geq E_0$ per unit comoving volume, will be used to fit the observed spectrum by the calculated one.

\begin{figure}
\begin{center}
\includegraphics[width=0.48\textwidth]{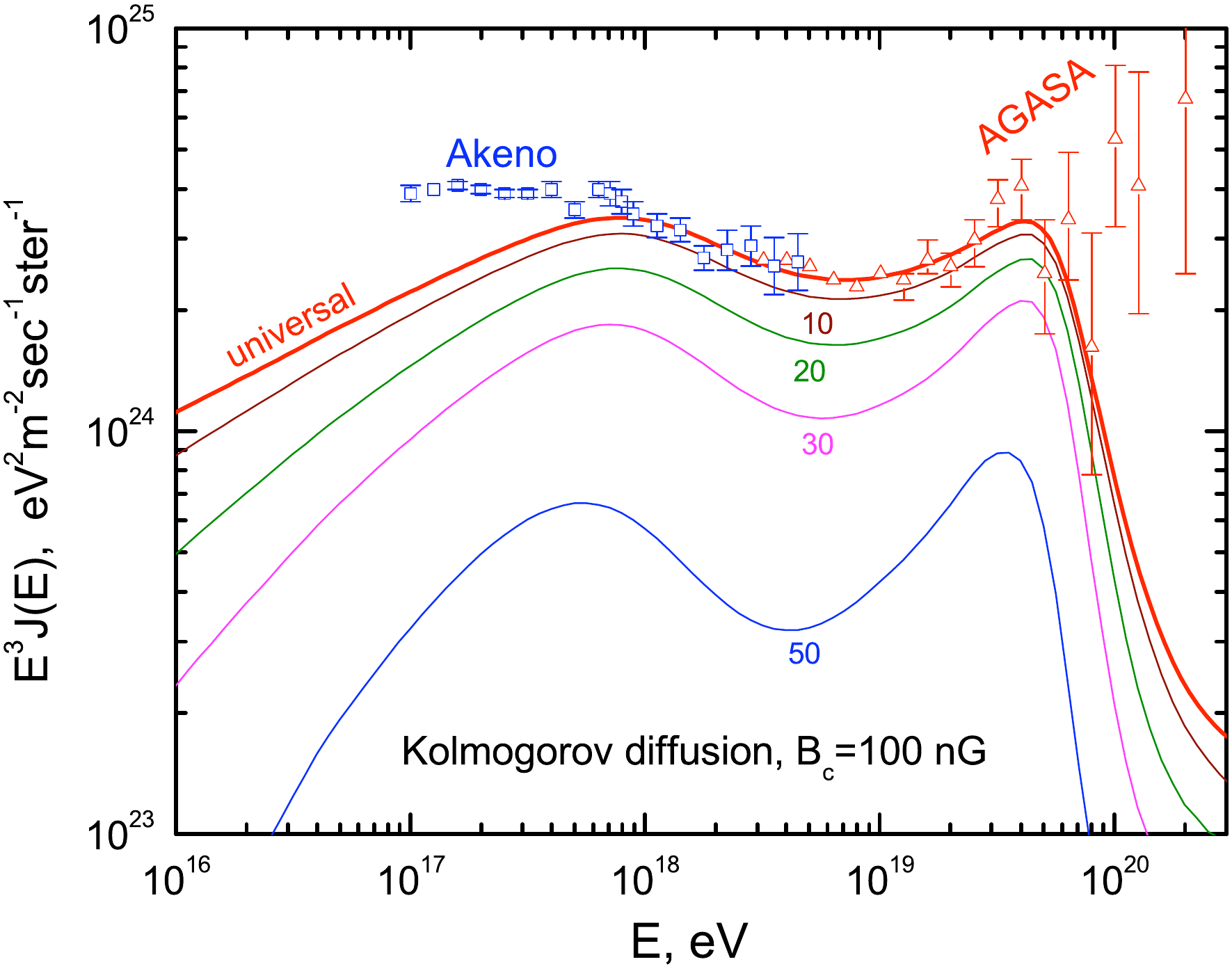}
\end{center}
\caption{Convergence of the diffusive solution to the
universal spectrum when the distance between sources diminishes
from 50 to 10~Mpc shown by numbers on the curves. }
\label{fig1} 
\end{figure}

In figure \ref{fig1} we test the BG solution with the help of the diffusion theorem
\cite{AB04}, which states that the diffusive solution converges to the universal spectrum, 
i.e. the flux computed with rectilinear propagation for an homogeneous distribution of sources, 
in the limit $d\to 0$, being $d$ the lattice spacing. Figure \ref{fig1} clearly shows this convergence even in the case of a strong magnetic field $B_c=100$ nG (and Kolmogorov 
diffusion).

\begin{figure}
\begin{center}
\includegraphics[width=0.48\textwidth]{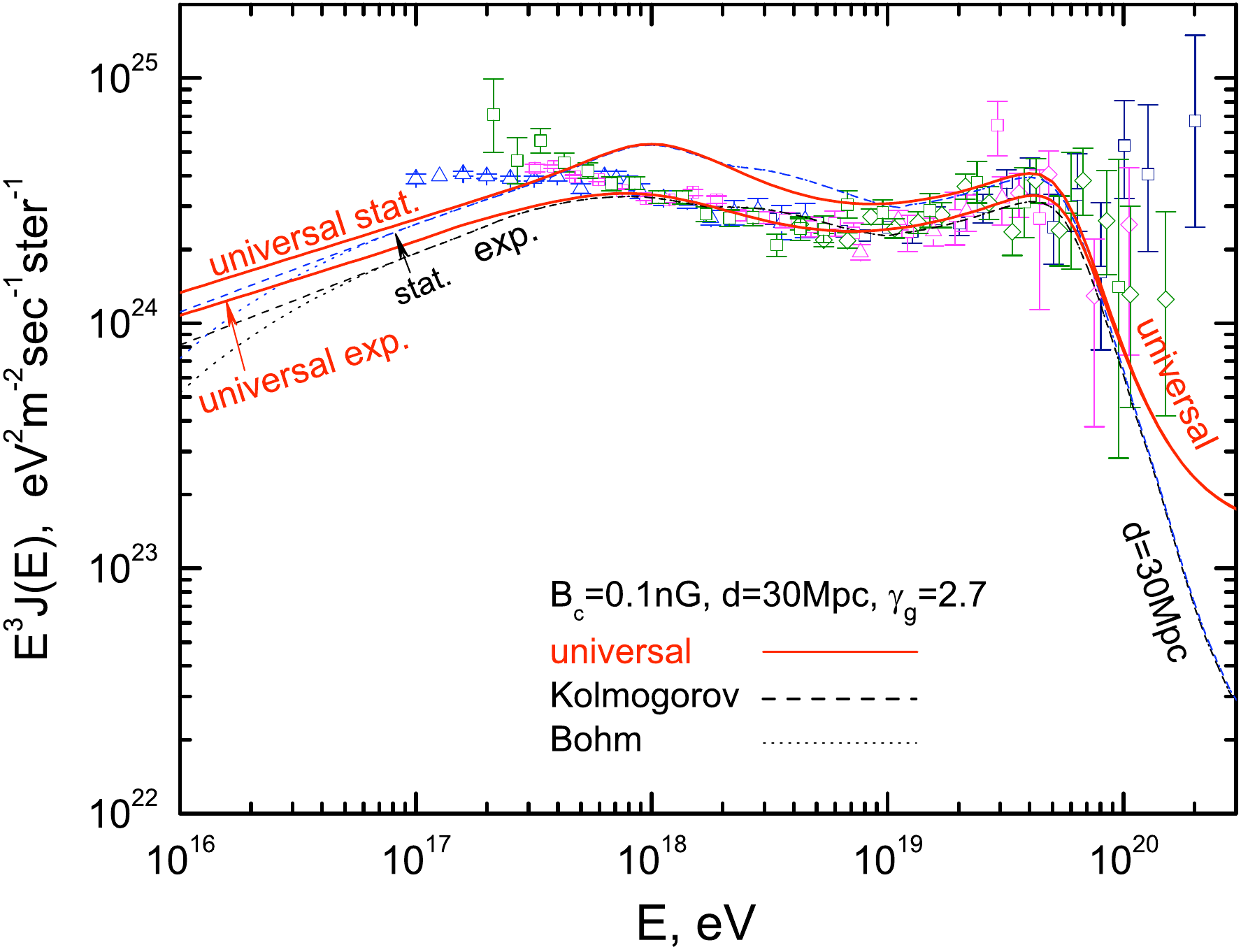}
\end{center}
\caption{Equal parameter comparison of the BG (expanding universe) and Syrovatsky (static universe) solutions, for $\gamma_g=2.7$, $\mathcal{L}_0 = 2.4 \times 10^{45}$~erg/Mpc$^3$yr and $d =  30$~Mpc. The magnetic field configuration assumed is $B_c=0.1$ nG and $l_c=1$ Mpc with 
different diffusion regimes as indicated on the plot. }
\label{fig2} 
\end{figure}

In the case of a small distance between source and observer the diffusive approximation 
is not valid. This result follows from a simple argument, the diffusive approximation works 
if the diffusive propagation time $r^2/D$ is larger than the time of rectilinear propagation, $r/c$. 
This condition, using $D \sim c \, l_d$, results in $r \geq l_d$. At distances $r \leq l_d$ the rectilinear and diffusive trajectories in IMF differ by a little quantity and rectilinear propagation is a good approximation as far as spectra are concerned. The number densities of particles  $Q/4\pi c r^2$ and
$Q/4\pi D r$, calculated in rectilinear and diffusive approximations, respectively, are equal at $r \sim l_d$, where $Q$ is the rate of particle production. We calculated the number densities of protons $n(E,r)$ numerically for both modes of propagations with energy losses of protons taken into account, and the transition is taken from the equality of the two spectra. We
know that this recipe is somewhat rough and an interpolation between the two regimes is required
\cite{AB05}. However, this interpolation is somewhat difficult because the diffusive regime sets up at distances not less than six diffusion lengths $l_d$. At distances $l_d \leq r \leq 6\, l_d$ some intermediate regime of propagation is valid. When studied in numerical simulations (e.g. 
\cite{Sato03}), the calculated number density $n(E,r)$ satisfies the particle number conservation $4\pi r^2 n u = Q$, where $u$ is the streaming velocity, while with a simple interpolated spectrum
this condition is not fulfilled a priori. In the present paper we will not address this problem, that will be
studied in a forthcoming paper \cite{ABG07}, assuming the rough recipe for the transition between 
diffusive and rectilinear regimes depicted above. This computation scheme can produce artificial features in the spectra, that are useful as a mark of the transition between the two regimes. 

\begin{figure}
\begin{center}
\includegraphics[width=0.48\textwidth]{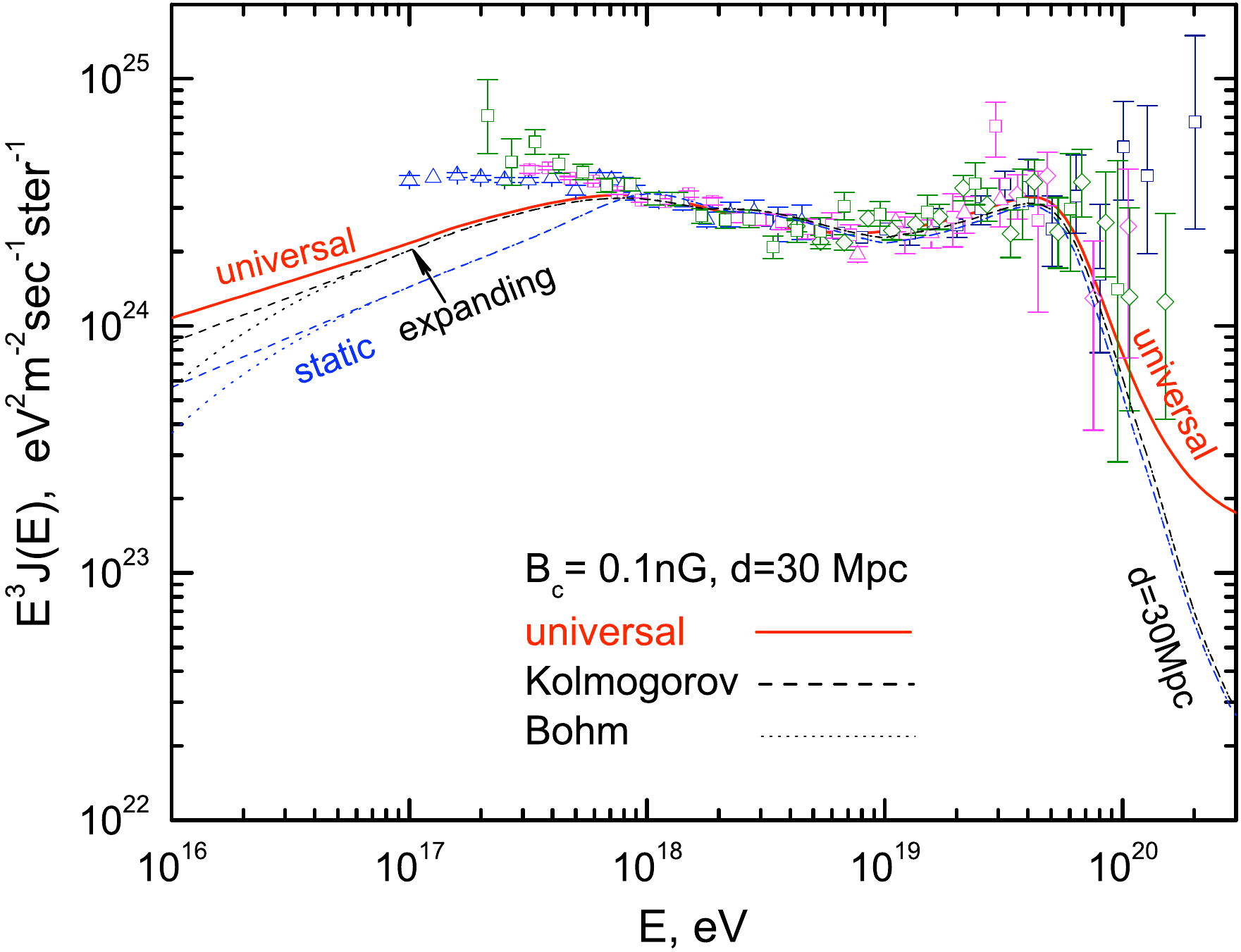}
\end{center}
\caption{Best fit comparison of the BG (expanding universe) and Syrovatsky (static universe) solutions, for $\gamma_g=2.7$, $\mathcal{L}_0 = 2.4 \times 10^{45}$~erg/Mpc$^3$yr and $d =  30$~Mpc. The magnetic field configuration assumed is $B_c=0.1$ nG and $l_c=1$ Mpc with 
different diffusion regimes as indicated on the plot. }
\label{fig3} 
\end{figure}

The direct comparison of the BG and Syrovatsky solutions of the diffusion equations is not possible because they are embedded in different cosmological environments. While the BG solution is
valid for an expanding universe, the Syrovatsky solution is valid only for a static universe. Using two different cosmological models for these solutions, there are two ways of comparison. The first one is given by equal values of parameters in both solutions. In this method for BG solution we use the standard cosmological parameters for an expanding universe $H_0$, $\Omega_m$, $\Lambda$ and maximum red-shift $z_{max}$ up to which UHECR sources are still active, magnetic field configuration ($B_c,l_c$), separation $d$ and UHECR parameters $\gamma_g$ and
$\mathcal{L}_0$, determined by the best fit of the observed spectrum. For a static universe with Syrovatsky solution we use the same parameters $H_0$, $d$, ($B_c, l_c$), $\gamma_g$ and
$\mathcal{L}_0$. The maximum red-shift in the BG solution is fixed by the age of the universe which equals to $t_0 =H_0^{-1}$ in the static universe ($z_{max}=1.5$). This formal method of comparison will be referred to as "equal-parameter method". Physically a better justified comparison is given by the best fit method, in which $\gamma_g$ and $\mathcal{L}_0$ are chosen as the best fit parameters for both solutions independently. 

The comparison of the two solutions is given in Figures \ref{fig2} and \ref{fig3} in the case of 
$B_c=0.1$ nG and $l_c=1$ Mpc with a source spacing $d=30$ Mpc. From these figures one can see a reasonably good agreement between the Syrovatsky solution, embedded in a static universe model, with the BG solution for an expanding universe at energies $E > 1 \times 10^{18}$~eV, at smaller energies appears a noticeable discrepancy between the two solutions that is natural and  understandable as discussed in the introduction. We conclude stating that, from a physical point 
of view, the second method of comparison is more meaningful and it gives a substantial agreement of the spectra obtained in the two cases. 

%This is the reference to .bib file (Without .bib!)
\bibliography{icrc1024}

\begin{thebibliography}{10}

\bibitem{AB04}
R.~{Aloisio} and V.~{Berezinsky}.
\newblock {Diffusive propagation of UHECR and the propagation theorem}.
\newblock {\em Astrtophys.J.}, 612:900--913, 2004.

\bibitem{AB05}
R.~{Aloisio} and V.~{Berezinsky}.
\newblock {Anti-GZK effect in UHECR diffusive propagation}.
\newblock {\em Astrtophys.J.}, 625:249--255, 2005.

\bibitem{Aloisioetal}
R.~{Aloisio}, V.~{Berezinsky}, P.~{Blasi}, A.~{Gazizov}, S.~{Grigorieva}, and
  B.~{Hnatyk}.
\newblock {A dip in the UHECR spectrum and the transition from galactic to
  extragalactic cosmic rays}.
\newblock {\em Astrop.Phys.}, 27:76--91, 2007.

\bibitem{ABG07}
R.~{Aloisio}, V.~{Berezinsky}, and A.~{Gazizov}.
\newblock {In preparation}.

\bibitem{BG1}
V.~{Berezinsky} and A.~{Gazizov}.
\newblock {Diffusion of cosmic rays in expanding universe}.
\newblock {\em Astrophys.J.}, 643:8--13, 2006.

\bibitem{BG2}
V.~{Berezinsky} and A.~{Gazizov}.
\newblock {Diffusion of Cosmic Rays in the Expanding Universe. 2. Energy
  Spectra of Ultra-High Energy Cosmic Rays}.
\newblock {\em astro-ph/0702102}, 2007.

\bibitem{BBO99}
P.~{Blasi}, S.~{Burles}, and A.~{Olinto}.
\newblock {Cosmological magnetic fields limits in an inhomogeneous universe}.
\newblock {\em Astrophys.J.}, 514:L79--L82, 1999.

\bibitem{Lemoine}
L.~{Lemoine}.
\newblock {Extra-galactic magnetic fields and the second knee in the cosmic-ray
  spectrum}.
\newblock {\em Phys. Rev. D}, 71:083007, 2005.

\bibitem{Syrov}
S.I. {Syrovatskii}.
\newblock {The distribution of the relativistic electrons in the Galaxy and the
  spectrum of magnetic bremmsstrahlung radio emission}.
\newblock {\em Astron.Zh.}, 36:17, 1959.

\bibitem{Sato03}
H.~{Yoshiguchi}, S.~{Nagataki}, S.~{Tsubaki}, and K.~{Sato}.
\newblock {Small scale clustering in isotropic arrival distribution of
  ultra-high energy cosmic rays and implications for their source candidate}.
\newblock {\em Astroph.J.}, 586:1211, 2003.

\end{thebibliography}
%This in the bibtex style, is ok.
\bibliographystyle{plain}
\end{document}